
%
%

\input harvmac.tex
%
%
%
%
\ifx\answ\bigans
\else
\output={
  \almostshipout{\leftline{\vbox{\pagebody\makefootline}}}\advancepageno
}
\fi
%
%
%
\def\mayer{\vbox{\sl\centerline{Department of Physics 0319}%
\centerline{University of California, San Diego}
\centerline{9500 Gilman Drive}
\centerline{La Jolla, CA 92093-0319}}}
%
%

%
%
\def\UCSD#1#2{\noindent#1\hfill #2%
\bigskip\supereject\global\hsize=\hsbody%
\footline={\hss\tenrm\folio\hss}}
%
%
\def\abstract#1{\centerline{\bf Abstract}\nobreak\medskip\nobreak\par #1}
%
%
%
%
\edef\tfontsize{ scaled\magstep3}
 \tfontsize  \tfontsize
 \tfontsize \font\titlei=cmmi10 \tfontsize
\font\titleis=cmmi7 \tfontsize \font\titleiss=cmmi5 \tfontsize
\font\titlesy=cmsy10 \tfontsize \font\titlesys=cmsy7 \tfontsize
\font\titlesyss=cmsy5 \tfontsize  \tfontsize
\skewchar\titlei='177 \skewchar\titleis='177 \skewchar\titleiss='177
\skewchar\titlesy='60 \skewchar\titlesys='60 \skewchar\titlesyss='60
%
%
%
%
%
\def\inv{^{\raise.15ex\hbox{${\scriptscriptstyle -}$}\kern-.05em 1}}
\def\lbar{{\lower.35ex\hbox{$\mathchar'26$}\mkern-10mu\lambda}} 

%
%
%
%
\def\dsl{\,\raise.15ex\hbox{/}\mkern-13.5mu D} 
\def\delsl{\raise.15ex\hbox{/}\kern-.57em\partial}
\def\Ksl{\hbox{/\kern-.6000em\rm K}}
\def\Asl{\hbox{/\kern-.6500em \rm A}}
\def\Dsl{\hbox{/\kern-.6000em\rm D}} 
\def\Qsl{\hbox{/\kern-.6000em\rm Q}}
\def\gradsl{\hbox{/\kern-.6500em$\nabla$}}
%
%
\def\lspace{\ifx\answ\bigans{}\else\qquad\fi}
\def\lbspace{\ifx\answ\bigans{}\else\hskip-.2in\fi} 
%
%
\def\boxeqn#1{\vcenter{\vbox{\hrule\hbox{\vrule\kern3pt\vbox{\kern3pt
        \hbox{${\displaystyle #1}$}\kern3pt}\kern3pt\vrule}\hrule}}}
%
%
\def\mbox#1#2{\vcenter{\hrule \hbox{\vrule height#2in
\kern#1in \vrule} \hrule}}
%
%
%
%

  \def\CO{{\cal O}}

%
%
%
%
%

%

\def\bar#1{\overline{#1}}
\def\vev#1{\left\langle #1 \right\rangle}

\def\abs#1{\left| #1\right|}

\def\darr#1{\raise1.5ex\hbox{$\leftrightarrow$}\mkern-16.5mu #1}

%
%
\def\frac#1#2{{\textstyle{#1\over #2}}} 
%
%
%
%

\def\Tr{\mathop{\rm Tr}}

%
%
%
%

%
%
\def\ltap{\ \raise.3ex\hbox{$<$\kern-.75em\lower1ex\hbox{$\sim$}}\ }
\def\gtap{\ \raise.3ex\hbox{$>$\kern-.75em\lower1ex\hbox{$\sim$}}\ }
\def\gl{\ \raise.5ex\hbox{$>$}\kern-.8em\lower.5ex\hbox{$<$}\ }
\def\roughly#1{\raise.3ex\hbox{$#1$\kern-.75em\lower1ex\hbox{$\sim$}}}
%
%
\def\ie{\hbox{\it i.e.}}        
        
\def\etal{\hbox{\it et al.}}

\def\np#1#2#3{{Nucl. Phys. } B{#1} (#2) #3}
\def\pl#1#2#3{{Phys. Lett. } {#1}B (#2) #3}
\def\prl#1#2#3{{Phys. Rev. Lett. } {#1} (#2) #3}
\def\physrev#1#2#3{{Phys. Rev. } {#1} (#2) #3}

\relax

\def\CO{{\cal O}}

\def\lta{\ \hbox{\raise.55ex\hbox{$<$}} \!\!\!\!\!
\hbox{\raise-.5ex\hbox{$\sim$}}\ }
\def\gta{\ \hbox{\raise.55ex\hbox{$>$}} \!\!\!\!\!
\hbox{\raise-.5ex\hbox{$\sim$}}\ }
\def\mayer{\vbox{\sl\centerline{Department of Physics}
\centerline{9500 Gilman Drive 0319}
\centerline{University of California, San Diego}
\centerline{La Jolla, CA 92093-0319}}}
\input epsf
\def\frac#1#2{{\textstyle{#1 \over #2}}}

\def\[{\left[}
\def\]{\right]}
\def\({\left(}
\def\){\right)}

\def\lfm{\medskip\noindent}
\def\inos{{\it inos }}
\def\ino{{\it ino }}
\noblackbox
\vskip 1.in

\centerline{{\titlefont{Supersymmetric      Baryogenesis}}}
 \medskip
\centerline{{\bf Submitted to Physics Letters B}}
\bigskip
\medskip
\centerline{A. G. Cohen\footnote{}{Email:
cohen@andy.bu.edu,  anelson@ucsd.edu}
\footnote{$^a$}{DOE Outstanding Junior Investigator}}
\centerline{{\sl
Physics Department}} \centerline{{\sl Boston University}}
\centerline{{\sl Boston, MA 02215}}
 \medskip
\centerline{  A. E.
Nelson\footnote{$^b$}{Sloan fellow, SSC fellow}  }
\bigskip\mayer \bigskip \vfill
\abstract{Requiring that the baryon number of the universe be generated by
anomalous electroweak interactions places strong constraints on the minimal
supersymmetric standard model. In particular,      the
electric dipole moment of the neutron must be greater than
$10^{-27}$e-cm.
Improvement of the current experimental bound on the neutron's electric dipole
moment  by one order of magnitude would constrain the lightest chargino to be
lighter than 88 GeV, and the  the lightest neutralino to be lighter than
44 GeV.   In    extensions of this model with gauge singlet superfields all of
these bounds
are eliminated.  } \vfill\UCSD{\hbox{UCSD/PTH
92-32,\
\break BU-HEP-92-20}}{August 1992}

Despite the success of the standard model of weak and strong
interactions, we still remain ignorant of the mechanism of electroweak
symmetry breaking; we are just beginning to probe this sector directly
through Higgs searches at LEP.   We are also still ignorant of
the origins of the CP violation observed in the kaon system.
Unfortunately, the minimal standard
model accounts for all solid experimental results observed to date, and we
have
few experimental   constraints on the symmetry breaking or CP violating
aspects of
the theory.  There is a strong cosmological argument that the minimal
standard
model cannot be the whole particle physics story. The baryon to
entropy ratio of the universe is $(0.4 - 1.0)10^{-10}$ \ref\kolbturner{E.
Kolb and
M. Turner, The Early Universe (Addison-Wesley, New York, (1990)}, and
explaining this
observable requires  baryon number violation coupled with out-of-equilibrium
CP
violation  in the early universe  \ref\barcon{A.D. Sakharov, JETP Lett. 6
(1967)
24}. While the standard model does have CP violation, the effects of the CP
violating phase in the Kobayashi-Maskawa matrix are too suppressed
by small masses and mixing angles
in order to account for the observed baryonic asymmetry of the universe (BAU)
\foot{Shaposhnikov has suggested two conceivable ways to enhance
the
CP violation in the standard model at high temperature  \ref\shaposh{M.E.
Shaposhnikov,  \np{287}{1987}{757}; \np{299}{1988}{797};
Phys. Lett. 277B (1992) 324,  Erratum, Phys. Lett. 282B
(1992) 483 }    ; the first
mechanism, dynamical high temperature spontaneous CP violation, is
contradicted by
non-perturbative computation \ref\afs{J. Ambjorn, K. Farakos, and M.E.
Shaposhnikov,
Niels Bohr Institute
preprint NBI-92-20 (1992)}, and the second mechanism, reflection of baryon
number
from expanding bubble walls,  according to our estimates cannot provide a
large
enough asymmetry \ref\shaposhaerr{Shaposhnikov's original estimate of the
BAU left
out several important suppression factors, such as cancellations in CP
violating
phases, the   slow rate of anomalous baryon number violation, and the large
width
of the bubble walls. See the Erratum in \shaposh. }.}. The standard model
also
contains anomalous baryon number violating interactions \ref\thooft{G.
t'Hooft,
\prl{37}{1976}{8}; \physrev{14}{1976}{3432}}, which should be  rapid enough
at
high temperatures   to affect cosmology \nref\early{A. Linde,
\pl{70}{1977}{306};
S. Dimopoulos and L. Susskind, \physrev{D18}{1978}{4500};
 N. Christ, Phys. Rev. D21 (1980) 1591; N.S. Manton, Phys.
Rev. D28 (1983) 2019; F.R. Klinkhammer and N.S. Manton, Phys. Rev. D30
(1984) 2212}\nref\krs{V.A. Kuzmin, V.A.
Rubakov and M.E. Shaposhnikov, \pl{155}{1985}{36} }\nref\more{P. Arnold and
L.
McLerran,   Phys. Rev. D36 (1987) 581;
   Phys. Rev. D37 (1988)
1020}\refs{\early-\more}.
Furthermore, the standard model can satisfy the
out of
equilibrium condition for baryogenesis if the phase transition is   first
order, proceeding via nucleation and expansion of bubbles of the broken
phase
\krs. Unfortunately it is necessary for  the vev of the Higgs field
after the transition to be large
 in order to avoid washing out any baryon number created
during the
transition \nref\higgslimit{M.E. Shaposhnikov, JETP Lett. {\bf 44}
(1986) 465;
A.I. Bochkarev, S. Yu. Khlebnikov and M.E. Shaposhnikov,
Nucl.~Phys. {\bf B329} (1990) 490.}\nref\dhs{M. Dine, P. Huet, and R.
Singleton, \np{375}{1992}{625}}\nref\wash{M. Dine, R. G.
Leigh, P. Huet, A. Linde, and D. Linde, SLAC preprint SLAC-PUB-5741 (1992);
B.H.
Liu, L. McLerran, and N. Turok, Minnesota preprint TPI-MINN-92/18-T
(1992)}\nref\superwashone{G.F. Giudice, \physrev{D45}{1992}{3177}}
\nref\superwashtwo{S. Myint, \pl{287}{1992}{325},
and
also work in progress}\refs{\shaposh,\higgslimit-\superwashtwo}. In the
minimal
standard
model this requirement cannot   be satisfied unless the Higgs is lighter
than
$\sim35$ GeV  \wash, which conflicts with current experimental bounds. Thus
we
should look beyond the standard model in order to discover the origin of
the BAU.
However we may not need to look very far, as  possible  mechanisms for
baryogenesis have  been suggested in several reasonable  extensions such as
axion
models with additional light scalar doublets \ref\mclerran{ L. McLerran,
\prl{62}{1989}{1075}}, the singlet Majoron model \ref\usone{A.G. Cohen,
D.B.
Kaplan and A.E. Nelson, \pl{245}{1990}{561};  \np{349}{1991}{727}}, the two
Higgs
model \nref\mtsv{N. Turok and  J. Zadrozny,
\prl{65}{1990}{2331}; \np{358}{1991}{471}; L. McLerran, M. Shaposhnikov, N.
Turok
and M. Voloshin,
\pl{256}{1991}{451}}\nref\ustwo{A.G. Cohen, D.B. Kaplan and A.E. Nelson,
\pl{263}{1991}{86}}\nref\usthree{A.E. Nelson, D.B. Kaplan and A.G. Cohen,
\np{373}{1992}{453}}\refs{\mtsv-\usthree}, the supersymmetric standard
model
\refs{\ustwo,\usthree}, extended supersymmetric models \ref\dhss{M.
Dine,
P. Huet, R. Singleton and L. Susskind,
\pl{257}{1991}{351}} and left-right symmetric
models \ref\moh{R.N. Mohapatra and
X. Zhang, Maryland preprint UMDHEP 92-230 (1992)}. Requiring that
sufficient BAU be
generated during the weak transition, and that it not be washed out afterwards
can give us new information about the CP violating and symmetry
breaking
sectors of the weak interactions,  allowing us to rule out some models
(such as
the minimal standard model), and to   constrain parameters in
others. It is
the aim of this letter to use baryogenesis to find new constraints on
supersymmetric models. We  are able to severely constrain
almost all
unknown parameters of the minimal model, while considerable freedom remains
for
models with additional gauge singlets.

First let us consider baryogenesis in the minimal supersymmetric
standard model (MSSM) \ref\mssm{S. Dimopoulos and H. Georgi,
\np{193}{1981}{150}; N. Sakai, Z. Phys. C11 (1981) 153}. Several authors
have
claimed that this model is ruled out   for baryogenesis because
the bound on the lightest scalar mass in this model is the same as in the
minimal
standard model   \refs{\dhs, \superwashone}. However
   Myint  \superwashtwo\ finds that the bounds
on the
scalar mass are relaxed somewhat, to 64 GeV, due to top  quark and squark
corrections to the high temperature effective potential when the top quark is
heavy and the
squark masses are not too heavy. The     bound is relaxed
because top
squarks can play the role envisioned by Anderson and Hall for a gauge
singlet
scalar, whose coupling to the Higgs doublet increases the strength of the
transition \ref\andhall{G. Anderson and L. Hall,
\physrev{D45}{1992}{2685}}. The
upper bound on the Higgs mass was computed in   one loop perturbation theory
and higher order corrections to the gauge propagator will reduce this number
by a factor of about $\sqrt{2/3}$ \wash, to $\sim 50$ GeV.
Furthermore this bound  will receive corrections  proportional to
$(m_H/m_Z)^2$
from
two loop corrections. Thus avoiding baryon
number washout is difficult in the MSSM, but not impossible. To increase
the
upper bound on the Higgs mass as much as possible, one should take    the
top
quark to be  heavy ( $ >  150$ GeV), the squark masses  not much heavier
than   150
GeV, the trilinear soft supersymmetry breaking terms (``A-terms'')   small,
and
the parameter $\tan\beta$ (the ratio of the two Higgs vevs)   less than
1.7.  Then, in improved one loop perturbation theory, avoiding baryon number
washout
requires   the lightest scalar mass to be lighter than  $\sim 50$ GeV,
which is
not in conflict with   current   bounds  provided that   higher order
corrections do not decrease the mass bound.

The MSSM has two
possible sources of CP violation which are absent in the minimal standard
model
\ref\ssmcp{J. Ellis, S. Ferrara, and D.V. Nanopoulos, \pl{114}{1982}{231};
W. Buchmuller and D. Wyler, \pl{121}{1983}{321}; J.
Polchinski and M.B. Wise, \pl{125}{1983}{393};
F. del Aguila, M. Gavela, J. Grifols, and A. Mendez, \pl{126}{1983}{71};
D.V. Nanopoulos and M. Srednicki, \pl{128}{1983}{61};
M. Dugan, B.
Grinstein and L.J.
Hall, \np{255}{1985}{413}}; a combination of these will be constrained
by the BAU.    The
interactions in this model are given by  \eqn\mssm{\eqalign{
&\left[\bar{U}\lambda_{U}QH
+\bar{D}\lambda_DQH'+\bar{E}\lambda_ELH'+\abs{\mu}e^{-i\phi_B}HH' \right]_F
\cr &
 +m_{3/2} \left[ \abs{A}e^{i\phi_A}(\bar{U}\xi_UQH
+\bar{D}\xi_DQH'+\bar{E}\xi_ELH')
+\abs{\mu_B}HH' \right]_A \ .\cr
}}
Note that
unlike the generic  two Higgs doublet model discussed in
refs.~\refs{\mtsv-\usthree}, the Higgs
potential does not contain
any CP violating phases at tree level---the phases $\phi_A$ and $\phi_B$
occur
only in interactions involving  the super-partners of the ordinary
particles. We find that the contribution of the phase $\phi_A$ to the BAU
is small, and so  we focus  on the
effects of the phase
$\phi_B$, which appears in the mass matrices of the
supersymmetric partners of the
 gauge bosons and the   Higgs scalars (the  ``{\it inos}'').

Our strategy for calculating the BAU in the  MSSM is as follows. We
will fix the top mass at 170 GeV, the lightest scalar mass at 48 GeV, the
soft
supersymmetry breaking  top squark masses at 150 GeV, the A-terms at zero,
and
$\beta$ at 0.85 ($\tan\beta=1.14$), since these parameters are already
constrained to be near these values \superwashtwo, and  in any case allowing
them to vary   will only affect the BAU by $\CO(1)$. Using these values we then
calculate the critical temperature, the effective potential at the critical
temperature, and the shape of the bubble walls. (Note that these phase
transition
quantities are not very sensitive to other supersymmetric parameters such
as
\ino masses.)    We use  the improved
one loop approximation for the effective potential,  including the order
$T^2$
corrections to the gauge propagators, and neglecting the contribution of
Higgs doublet
loops. When the mass of the lightest scalar is  far below the gauge
boson
masses this approximation is reliable at the critical temperature around
the symmetry breaking minimum and   in the vicinity of the symmetric
minimum for
scalar field expectation values larger than $\CO(g_{\rm wk}T)$. Although
perturbation theory is not valid for a  calculation of the   effective
potential
between
the two minima, which will affect the width and shape of the phase
boundary,
fortunately our calculation of the BAU will turn out to be insensitive to the
detailed
shape of this boundary, provided it is much thicker than $\CO(1)/T$.
The one
loop estimate   gives     the width of the phase boundary to be $11/T$ at
the
temperature where the two minima are degenerate, so we will assume
the phase boundary is thick. The one loop effective potential indicates
that the
ratio of the   expectation values of the two Higgs remains constant during
the
transition, so   we take
\eqn\rat{
H'=H\({\vev{H'}\over\vev{H}}\) ,}where $\vev{H'}$ and $\vev{H}$   are the
expectation values at the critical temperature in the symmetry breaking
minimum. The   temperature of the transition\foot{ Our definition
of the
transition temperature is the spinodal point where the local minimum at the
origin
vanishes.
This temperature is slightly lower than the  one loop estimates of the
temperature
at
which the transition actually occurs \wash, but these estimates require
knowing
the effective potential in a region where it is not calculable
perturbatively.
We use
this definition in order to get a conservative upper bound on the BAU
produced
during
the transition.} is 59 GeV, and at this temperature the minimum of the
effective
potential
occurs at $H=63$ GeV and $H'=53$ GeV.

For     thick   bubble walls     the
relevant
baryogenesis mechanism is known as ``spontaneous baryogenesis'', reviewed
below.
This mechanism, first
introduced for  baryogenesis during a second order phase transition
\ref\ck{A.G. Cohen and D.B. Kaplan, \pl{199}{1988}{251};
\np{308}{1988}{913}},
involves  a space-time dependent
 field, which evolves coherently during the transition. This  time
evolution produces a  CPT-violating term  in the effective Hamiltonian
called a ``charge potential'', which resembles a  chemical
potential\foot{Recall
that a chemical potential is a Lagrange multiplier introduced to implement
a
constraint, which appears in the effective Hamiltonian as an energy
splitting
between particles and anti-particles. Similarly, a  charge potential results
in
different energy levels for particles and anti-particles, but the energy
difference
is a real physical effect resulting from dynamical violation of CPT during
a phase
transition.}. A charge potential will cause  the  free energy density
inside the
bubble walls to be minimized  for  nonzero baryon number, and hence the
production of a
net baryon number via anomalous weak interactions. Spontaneous electroweak
baryogenesis during the weak phase transition  has been  suggested   for
the  two Higgs model in refs.~\refs{\mtsv,\ustwo}  and   as a
baryogenesis mechanism for supersymmetric  models in refs.~\refs{\dhss,\ustwo}.

   In the
MSSM
the \ino mass matrices are space-time varying during the transition, and
contain an
irremovable CP violating phase, which leads to a  charge
potential
for baryon number in the effective
fermion Hamiltonian.   If the \inos are not too heavy and the phase is not
too
small, this charge potential  is large enough to result  in generation of
an
acceptable baryon number during the weak transition.
We will find that the resultant BAU depends sensitively on the
CP violating phase $\phi_B$,
and on
the masses of the \inos. We are
able to
use the BAU to
place upper bounds on \ino masses and   lower bounds on $\phi_B$ and the
electric
dipole moment of the neutron.

Before launching into the specifics of the MSSM calculation,  we show how
to
calculate the charge potentials resulting from space-time varying fermion
mass
matrices.

Consider a fermion mass term of the form
\eqn\massterm{ \psi_{i }^T C m_{ij}(x_\mu)\psi_{j }+\hbox{h.c.}\ ,}
where  we take all fermions to be left-handed, and $C$ is the charge
conjugation
matrix .  We
can   make a space-time
dependent
unitary change of basis on the fermions: \eqn\chngebasis{ \psi_{i
}\rightarrow
U_{ij}(x_\mu)\psi_j \ . }  in order to make  the
fermion mass terms  everywhere real, positive and diagonal; however the
space-time dependence of $U$ requires that we replace the
kinetic
energy terms in the Lagrangian by
\eqn\lkin{
{\cal L}_{\hbox{K.E.}} \rightarrow {\cal L}_{\hbox{K.E.}}
 + \bar\psi \gamma^\mu (U^\dagger  i\partial_\mu U) \psi
   \ .}
 Note that since $U$ is a unitary matrix,
$U^\dagger\partial_\mu U$ may be written
\eqn\gener{U^\dagger\partial_\mu
U= i\partial_\mu \sum_a \alpha_a(x_\mu) T_a \ , }
where for $n$ fermions the
$T_a$s are generators of $U(n)$, and the functions $\alpha(x_\mu)$ are
defined by
eq.~\gener. Thus the Lagrangian with mass term \massterm\ is equivalent to
a
Lagrangian with a real diagonal mass term but also containing a term
\eqn\chargepot{ - \sum_a \partial_\mu\alpha_a(x_\mu)\bar\psi
 T_a\gamma^\mu\psi
  \ .}
  If the transformation
\chngebasis\ has a gauge anomaly there will also be a modification of the
Lagrangian \eqn\anom{\sum_\beta\theta F_\beta\tilde F_\beta\rightarrow
\sum_{\beta
a}\(\theta+ { g_\beta^2\over16\pi^2} \alpha_a\Tr T_a
 t_\beta^{2}   \) F_\beta\tilde F_\beta  ,} where the $t_\beta$s are
gauge generators in the left-handed fermion representation  and the $F_\beta$s
are the gauge field strengths.

The presence of these anomalous terms complicates the discussion of the
charge
potentials, and consequently we will choose our unitary transformation to
have no
gauge anomaly; this is the strategy followed in \ck.

Finally, if the
transformation~\chngebasis\ does not correspond to a symmetry of the
interactions
the coupling constants will be affected by the change of basis; however
the
effect of these coupling constants on energy levels is higher order in
perturbation
theory and will not concern us here.

    What effect does a charge potential have on
a thermal system? There are  two possibilities:
\lfm
\item{1)} If there is a charge potential for an exactly conserved charge,
(e.g.
electric charge or B-L) we can ignore it.
Although it looks like the system
could lower its free energy by producing a net charge, charge conservation
imposes a zero charge constraint.  It can easily be seen by integration by
parts that a charge potential for an exactly conserved charge has no physical
effect; equivalently, we can always redefine fields so such charge potentials
never arise.

\item{2)} Charge potentials for non-conserved charges  will
lead to an asymmetry in the rates between processes which create and
destroy the
charges, until the system reaches thermal equilibrium.   The thermal
occupation numbers in general will be different for
particles and their CP conjugates, due to CP violation and the dynamical
CPT
violation from the space-time varying scalar fields. For a   charge which
is
approximately (but not exactly) conserved  the system will take a long time
to reach
equilibrium. If the system is near thermal equilibrium  the net rate of
charge
production can be computed using thermodynamic arguments. For instance  if
there
is a small  charge potential $\dot\alpha_B$
for the baryon number current and no other charge potentials\foot{Note that
anomalous weak baryon number violating processes can be affected by a
potential
for any charge generator whose trace over left handed fermion weak doublets
is
nonzero~\ustwo.},  for a   system starting with  no net quantum numbers
the
constraint of zero net baryon number can be implemented by introducing a
baryon chemical potential \eqn\barchem{\mu_B=-\dot\alpha_B\ .} The chemical
potential $\mu_B$  is just the force of constraint on the system, {\it
i.e.}
the derivative of the free energy with respect to baryon number. Now the
difference between the rates of anomalous processes which create and
destroy
baryon number is just proportional to   the  difference in the change in
the free
energy per event, leading to an anomalous baryon creation rate of
\eqn\bdot{\dot\rho_B= -{9 \mu_B   \Gamma_B\over T}=
{9  \dot\alpha_B \Gamma_B\over T}\ ,}
where $\Gamma_B$ is the rate of anomalous baryon violating events per unit
volume.
The factor of 9 comes about because each anomalous event changes the free
energy
by $3 \mu_B$ and changes the baryon number by 3 units.
\lfm

We can now calculate the BAU produced in the MSSM during the weak phase
transition. We first compute the charged and neutral
\ino mass matrices, as functions of the scalar field expectation values,
assuming
equ.~\rat, and find the transformation on the \ino fields  $U_I(  H )$ which
renders
their masses
real, positive, and diagonal.  $U_I$  will in general be anomalous and
would, by itself, give rise to an effective $W\tilde{W}$ operator,
as well as giving rise to
charge potentials for the \inos via equ.~\lkin.
By further making a space-time dependent baryon number rotation of
the
light fermions we can remove this anomalous operator,
 at the cost of introducing a charge potential for baryon number\foot{This
could
also be seen simply by using the anomaly equation to replace $F\tilde F$
with
the divergence of the baryon current, and integrating by parts \dhss.}:
\eqn\sponbarop{  {1\over 3}\partial_\mu \( \int^{\vev{H(x)}}_0  d H \Tr
U_I^\dagger (H) {  idU_I(H)\over d H}
   t_{\rm wk}^2 \) j_B^\mu \  .
}
  We  then need to know how  anomalous processes  inside the bubble
walls are
affected by CP violating terms.
 Since the \inos can quickly   come into equilibrium with the
\ino charge potentials via ordinary {\it non-anomalous} interactions the \ino
charge
potentials will not affect anomalous processes\foot{In the limit of
vanishing
gaugino mass or $\mu$ there is an additional approximate symmetry and the
anomalous
baryon production rate will be further suppressed. We always assume we are
far
from this limit, \ie that the \ino masses are not much smaller than the
temperature. }. The main effect on anomalous baryogenesis will come from
the term
\sponbarop. If the system is near  thermal equilibrium then eqs. \bdot,
and
\sponbarop\ can be used to find  the total baryon number density
produced
during the transition\foot{The effect of the spatial component of the charge
potential will be to also produce a baryon number current inside the walls.
We find this current has no effect on the baryon density produced in the
thermal frame in the limit where the walls are thick. }
\eqn\totbarnum{\eqalign{
\rho_B=&\int dt \dot\rho_B\cr =&\int dt {9 \Gamma_B(\vev{H })\over T} {d\over
d
t}\({1\over 3} \int^{\vev{H }}_0  d H \Tr  U_I^\dagger (H){idU_I(H)\over d H}
   t_{\rm wk}^2 \)\cr
=&{3\over T}\int^{\vev{H }}_0  d H  \Tr U_I^\dagger (H){ idU_I(H)\over d H}
   t_{\rm wk}^2  \Gamma_B( H )  \ . \cr}}
We still need  to know the rate of anomalous baryon  density
production $\Gamma_B$, which has been estimated in the symmetric phase
to  be
\ref\estkappa{J. Ambjorn, T. Askgaard, H. Porter and M. Shaposhnikov, Phys.
Lett.
244B (1990) 479 }
 \eqn\aaps{\Gamma_B\sim  \alpha_{\rm wk}^4 T^4\ ,   }
while in the broken phase it is vanishingly small.
Unfortunately,   there is currently no reliable way of computing
$\Gamma_B$ inside the wall where the scalar expectation values are
changing.
Furthermore it has been claimed, based on some $1+1$ dimensional
simulations
\ref\gst{D. Grigoriev, M. Shaposhnikov and N. Turok, Princeton preprint
PUPT-91-1275 (1991)} that anomalous baryon production inside the wall is an
inherently non-equilibrium process. (Dine suggests this is unlikely to be the
case for a
 transition with thick walls and small latent heat such as occurs in the
MSSM~\ref\dinetalk{M.
Dine, talk given at the Yale-TEXAS symposium on electroweak baryon
violation,
(March 1992) SCIPP 92/21}.)  We will use a
conservative estimate for the maximum baryon number   produced
by
simply  computing the integral \totbarnum\ using equ.~\aaps\ for
$\Gamma_B$.  We think it is more likely that the baryon production inside the
wall will be suppressed when the Higgs vevs become large. For instance
McLerran
\ref\mclerrantwo{L. McLerran, talk  given at the
ITP workshop on Cosmological Phase transitions, Santa Barbara, April, 1992 }
estimates that baryon production is shut off when the Higgs vev reaches about
(1/2) its value inside the wall, while Dine \etal \refs{\dhs,\dhss} estimate
that this shutoff occurs for vevs of order $g_{\rm wk} T/(4 \pi)$. In Figure 1
we plot the baryon number produced as a function of the value of the Higgs
vev where this shutoff occurs for a typical choice of parameters; note that
the dependence is approximately quadratic. We conclude that it is most likely
that the actual baryon number produced during the transition will   be between
$\sim 1/4$  and $\sim 10^{-2}$ times smaller than our most favorable estimate.


In order to find the allowed range of parameters in the MSSM,
for each value of the gaugino and higgsino masses we adjust the CP
violating
phase $\phi_B$ to be as large as is compatible with current
constraints on the EDMN \ref\susyedmn{R. Arnowitt, J.L. Lopez, and D.V.
Nanopoulos,
\physrev{D42}{1990}{2423}; Y. Kizukuri and N. Oshimo,
\physrev{D45}{1992}{1806}}. We
then compute the  BAU, from equs.~\totbarnum\ and
\aaps. The allowed range of \ino masses are those for which the upper
bound on
the baryon to entropy ratio is greater than $0.4\times 10^{-10}.$ For
masses in the
allowed range, we compute the minimum value of $\phi_B$ which is consistent
with the
BAU, and use the calculation of Kizukuri and Oshimo in ref.~\susyedmn\ to
find the
lower bound on the EDMN.

In Figures~2a  and 2b
we plot the lower bound on the EDMN, as a function of $\mu$
and the wino
mass parameter $m_2$. The first figure has a phase $\phi_B$ near $0$,
while the second figure has the phase near $\pi$.
The bino mass parameter $m_1$ is assumed to satisfy
the GUT
relation $m_1=m_2(5/3)\tan^2\theta_W$. The signs in fig.~2b indicate the sign
of
the EDMN, while in fig.~2a the EDMN is negative\foot{In computing the EDMN, we
use  the quark model and   neglect  the contribution of the phase
$\phi_A$; we have assumed  the A-terms are small.}.
The black excluded region corresponds to an EDMN greater than the current bound
of $10^{-25}$ e-cm \ref\edmnbound{K. Smith \etal., \pl{234}{1990}{191}}, the
dark
grey region is an EDMN greater than $10^{-26}$ e-cm, and the light grey region
greater than $10^{-27}$ e-cm. Thus for any portion of the allowed parameter
space the EDMN is greater than $10^{-27}$ e-cm. (Similarly, if we assume the
selectron mass is equal to the
squark mass, the electric dipole moment of the electron is greater than
$\CO(3) \times 10^{-29}$ e-cm.) Throughout all of the region where the EDMN
could be less than
 $10^{-26}$ e-cm the lightest chargino mass is lighter than 88 GeV
and the
lightest neutralino mass is lighter than 44 GeV. Also, if the EDMN is less than
$10^{-26}$ e-cm the phase $\phi_B$ must be near $\pi$ in order to avoid
a chargino mass lighter than $M_Z/2$.


These figures have assumed that the rate of baryon violation is given by
equ.~\estkappa, and that this violation is occurring throughout the bubble
wall.
If this is not the case, the baryon number produced will be reduced and the
resulting phase must then be larger to compensate, giving rise to larger dipole
moments.

If the experimental bound on the
EDMN is pushed down by an  order  of magnitude,   or if numerical calculation
finds the rate of anomalous
baryon creation inside the phase boundary to be much less than it is in the
symmetric
phase, then there are strict upper bounds on
\ino masses.   If the MSSM
is
responsible for baryogenesis the prospects for discovering electric dipole
moments
and supersymmetric particles in the next few years are excellent.
Should we fail to make these discoveries, there are several possible
conclusions. The BAU may come from physics above the weak scale, from weak
scale physics other than supersymmetry, or from a more complicated
supersymmetric model. The simplest extension of the MSSM would be to add a
gauge singlet superfield, which changes the allowed form of the Higgs
potential
and greatly relaxes  all the constraints on scalar masses \ref\pietroni{M.
Pietroni,
Padova preprint PFPD/92/TH/36 (1992)}. The tree level potential  will
generically
contain large cubic terms, and   gives a very strongly first order
transition
with a thin boundary between the two phases. Furthermore, with gauge
singlets there
is   the possibility for CP violating phases in the Higgs potential. With
thin
bubble walls we expect baryogenesis to be dominated by the mechanism of
refs.~\refs{\usone,\usthree}, in which CP violating particle scattering
processes
from the phase boundary leads to a     transport of particle quantum
numbers, biasing anomalous baryon production throughout the symmetric
phase. The
charge transport mechanism could produce the BAU for phases as small as
$10^{-4}$
\usthree. Unfortunately such extended supersymmetric models currently have
too many
free parameters to allow for calculation of the BAU.

In summary, we have shown that unlike the minimal standard model, the MSSM
is still
viable, but soon   either it should be ruled out (for baryogenesis), or new
experimental discoveries such as electric dipole moments  will give us an
important
clue towards the  explanation of why there are more baryons than
anti-baryons.  We
are still a long way from testable predictions from baryogenesis in
extensions of the
MSSM.

 \bigbreak\bigskip\bigskip\centerline{\bf
Acknowledgments}\nobreak
We would like to thank David Kaplan and Stanley Myint for useful
discussions.
 A. C. was supported in part by DOE
contracts \#DE-AC02-89ER40509, \#DE-FG02-91ER40676,  by the Texas National
Research Laboratory Commission grant \#RGFY91B6,
and NSF contract \#PHY-9057173;
A. N.  was supported in part by DOE contract \#DE-FG03-90ER40546, by
a fellowship  from the Alfred P. Sloan Foundation, and by an SSC Fellowship
from
the Texas National Research Laboratory Commission.
 \bigbreak\bigskip\bigskip

\listrefs
\figures
\fig{1}{Baryon number density as a function of the Higgs vev.}
\fig{2a}{Lower bound on the EDMN  for phase $\phi_B$ near $0$. The
black region is excluded, the grey region has an EDMN
greater than $10^{-26}$ e-cm, and the white region has an
EDMN greater than $10^{-27}$ e-cm.}
\fig{2b}{Same as Figure 2a, except the phase $\phi_B$ is near $\pi$.
The $\pm$ signs indicate the sign of the contribution of $\phi_B$ to
the  down quark electric dipole moment at one loop, when $\phi_B$ is chosen
to produce a positive BAU.}

\bye